\documentclass[prb,twocolumn,showpacs,preprintnumbers,amsmath,amssymb]{revtex4}
\usepackage{dcolumn}
\usepackage{bm,amsmath,amssymb}
\usepackage{graphicx}
\usepackage{bm}
\begin{document}
\title{Changeover from Glassy ferromagnetism of the orbital domain state to long range ferromagnetic ordering in La${_{0.9}}$Sr$_{0.1}$MnO${_3}$}

\author{K. Mukherjee and A. Banerjee}
\affiliation{UGC-DAE Consortium for Scientific Research\\University Campus, Khandwa Road\\
Indore-452017, M.P, India.}
\date{\today}
\begin{abstract}
An attempt is made to resolve the controversy related to the low temperature phase (ground state) of the low-doped ferromagnetic (FM)- insulator(I) manganite through bulk magnetic measurements on La${_{0.9}}$Sr$_{0.1}$MnO${_3}$ sample. It is shown that the FM phase, formed out of well defined transition in the low-doped system, becomes inhomogeneous with decrease in temperature. This inhomogeniety is considered to be an outcome of the formation of orbital domain state of e$_g$-electrons having hole rich (metallic) walls separating the hole deficient (insulating) regions.  The resulting complexity brings in metastability and glassy behaviour within the FM phase at low temperature, however, with no resemblance to spin glass, cluster glass or reentrant phases. It shows ageing effect without memory but magnetic relaxation shows signatures of inter-cluster interaction. The energy landscape picture of this glassy phase is described in terms of hierarchical model. Further, it is shown that this inhomogeneity disappear in La${_{0.9}}$Sr$_{0.1}$MnO$_{3.08}$ where, the orbital domain state is destroyed by self doping resulting in reduction of Mn$^{3+}$ and hence e$_g$-electrons. The ferromagnetic phase of the non-stoichiometric sample,  does not show glassy behaviour. It neither follows 'hierarchical model' nor 'droplet model' generally used to explain glassy or inhomogeneous systems. Its magnetic response can be explained simply from the domain wall dynamics of otherwise homogeneous ferromagnet.   
\end{abstract}
\maketitle
\section {Introduction}
The physics controlling the properties of low-doped manganites is a subject of intense research currently, due to the fact that magnetic ground state of these compounds continues to be a subject of controversy\cite{kaw,yam,nij,end,joy,alg,pap,kaj,gec,choi}. The physical properties exhibited by these compounds are likely to be proximate to those of other low doped transition metal oxides, like cuprates and nickelates. Lightly doped manganites show ferromagnetic insulating behaviour inspite of finite amount of hole doping indicating that these transition metal oxides (TMO) belong to a class of strongly correlated electron system and the effect of correlation among the electrons prevents the ground state from being metallic. Introduction of holes in these systems results in inhomogeneity, which divides the system into different regions having varying hole densities. Generally in these systems the kinetic and potential energies are of the same energy scale and incorporation of coulomb interaction in these regimes leads to various self organised structures, with clusters of one phase embedded in the other, a phenomenon referred as electronic phase separation (EPS).\cite{mor} Recent theoritical studies also highlight the role of coulomb interaction in studying the electronic inhomogeneity in manganites.\cite{vij} EPS either results in formation of regions having competing magnetic interactions or in self-generated clusters, interaction among which results in blocking\cite{suni} or freezing mechanism (observed generally in TMO) at certain temperatures. Studies on cuprates\cite{zan,cho} and nickelates\cite{tra} have revealed a microscopic segregation of doped holes in antiferromagnetic phase into walls leading to an ordering consisting of charged domain walls that forms antiphase boundaries between antiferromagnetic domains. Studies by Tranquada et.al on La${_{1.48}}$Nd${_{0.4}}$Sr$_{0.4}$CuO${_4}$ \cite{tran} and La${_{1.69}}$Sr${_{0.31}}$NiO${_4}$\cite{tranq} have revealed stripe phase order of hole and spins. La${_{0.9}}$Sr$_{0.1}$MnO${_3}$ with x = 0.1 to 0.17 belong to the class of compound known as ferromagnetic insulator (FI). The self-organised regime observed in nickelates and cuprates is also expected in FI phase of manganite. Recent theories\cite{miz} and experiments\cite{choi} have provided evidences that orbital ordering (OO) plays an important role along with spin and charge in the insulating state of low doped manganites by controlling the e$_g$ electron mobility. Experimental results on La${_{0.88}}$Sr$_{0.12}$MnO${_3}$\cite{end} indicates the transition between two ferromagnetic phases, one metallic another insulating driven by orbital ordering. It has been proposed that the OO phase might contain ferromagnetic insulating domains separated by ferromagnetic metallic walls\cite{pap}, which raises question of stripes formation in the ferromagnetic insulating phase\cite{hot} 

Temperature dependence of ac susceptibility and low field magnetisation of low doped LaMnO${_3}$ shows interesting magnetic behaviour. Urushibara et al.\cite{uru} showed that La${_{0.9}}$Sr$_{0.1}$MnO${_3}$ is orthorhombic and undergoes a paramagnetic to ferromagnetic transition followed by a transition at low temperature accompanied with insulating behaviour. Susceptibility of ferromagnets generally varies as the inverse of demagnetisation factor and is expected to be constant at low temperature in absence of any further magnetic transition. Studies of the critical regimes in La${_{0.875}}$Sr$_{0.125}$MnO${_3}$ revealed that the paramagnetic to ferromagnetic phase transition is accompanied by consistent critical exponents belonging to 3D Heisenberg universility class.\cite{sunil} However, the low temperature transition which received considerable attention in the last decade remains undetermined. The nature of this low temperature phase previously has been interpreted in terms of canted antiferromagnetic phase\cite{kaw}. Successive structural phase transition from a high temperature pseudocubic phase to intermediate Jahn-Teller distorted orthorhombic phase and to low temperature pseudocubic phase reported in this low doped regime\cite{ark, kawa} might be a cause of the observed behaviour. The nature of low temperature state is also reported in terms of charge localization, which is accompanied by ordering of polarons\cite{yam}. A field induced phase transition from a ferromagnetic metallic phase to a ferromagnetic insulating phase as reported in La${_{0.875}}$Sr$_{0.125}$MnO${_3}$\cite{nij} might be responsible for the observed low temperature phenomenom. The low temperature fall in magnetisation in La${_{0.9}}$Ca$_{0.1}$MnO${_3}$ is intrepreted in terms of domain wall pinning effect by Joy et al.\cite{joy}. The two successive transitions with the lowering of temperature can be due to re-entrant spin glass transition.\cite{sat} Recent studies from neutron diffraction, small angle neutron scattering and nuclear magnetic resonance shows that the ground state of La${_{0.9}}$Ca$_{0.1}$MnO${_3}$ consists of disordered double exchange metallic clusters that co-exists with long range superexchange based ferromagnetic insulating regions.\cite{alg} Hence the above reports show a wide variety of possibilities of the low temperature phase and ground state of such low doped FI manganites.

In manganites the ratio of Mn$^{3+}$/Mn$^{4+}$ and their distribution in the lattice plays an important role in tuning the physical properties of these systems. Generally, the amount of Mn$^{4+}$ is tuned by divalent cation doping on A-site of perovskite structure but its amount can also be increased by oxidation of the stoichiometric sample resulting in a change of physical properties of the compound arising out of non-stoichiometry. The excess oxygen is accounted by  an equal number of vacancies at A and B sites of ABO$_3$ perovskites while the oxygen network in believed to be undefected.\cite{roo} As said earlier, investigations of low-doped manganites have revealed the formation of orbital domain state in the ferromagnetic insulating regimes.\cite{choi,pap} Increasing non stoichiometry, increases the Mn$^{4+}$ content resulting in suppression of the OO phase with the clusters becoming more populous eventually coalescing leading to the establishment of homogeneous ferromagnetic order. Hence the metamagnetic behaviour of the OO phase is expected to decrease with the increasing nonstoichiometry. Such stoichiometry dependent behaviour is also observed in other transition metal oxides like cuprates\cite{cho} and cobaltates\cite{sen} and more recently in bilayered manganite.\cite{qin}

In this paper, through bulk magnetization we investigate the magnetic ground state of stoichiometric La${_{0.9}}$Sr$_{0.1}$MnO${_3}$, as its true nature is in the centre of debate. A detailed investigation of the effect of both AC and DC magnetic field on the physical properties of this compound indicates that the observed low temperature behaviour is not because of magnetic transition but due to development of an inhomogeneous phase with the reduction in temperature.  The resulting self-organized regimes are of the form of orbital domains, dictated by OO, which plays an important role in defining the ground states of the compound. Increase of disorder (in form of self doping) suppresses the orbital domain state and a homogeneous ferromagnetic ordering is observed. The magnetic behaviour of the resulting non stoichiometric compound is ascribed to domain wall dynamics in a ferromagnetic matrix whereas, the orbital domains of the stoichiometric sample show glassy ferromagnetic behaviour with the glassiness arising solely due to intercluster interaction. The results are also compared with the hierarchical model.  
\maketitle\section{Sample preparation and characterization}
Two polycrystalline sample La${_{0.9}}$Sr$_{0.1}$MnO${_3}$ (S1) and La${_{0.9}}$Sr$_{0.1}$MnO$_{3.08}$ (S2) has been prepared by standard solid-state ceramic route with starting materials having purity ($>99.9\%$). Stoichiometric proportions of the starting materials La${_2}$O${_3}$, Sr${_2}$CO${_3}$ and MnO${_3}$ were mixed and heated in air at 950$^0$C  for 24hrs twice. After grinding, the powder, pellets were made and given a heat treatment of 1250$^0$C. For sample S1 the final sintering temperature is given at 1400$^0$C. Then it is annealed under nitrogen atmosphere for 36 hours. The sample S2 is annealed under oxygen atmosphere for 24 hours after the final sintering temperature of 1250$^0$C. X-Ray diffraction (XRD) was carried out using Rigaku Rotaflex RTC 300 RC diffractometer with CuKa radiation. The collected XRD pattern is analysed by the Rietveld profile refinement, using the profile refinement program by Young et al.\cite{you}. Estimation of Mn$^{3+}$/Mn$^{4+}$ is done by iodometric redox titration using sodium thiosulphate and potassium iodide. AC susceptibility and dc magnetisation is done using a homemade ac susceptometer\cite{ash} and a vibrating sample magnetometer (VSM)\cite{krs}. 
 
The samples S1 and S2 crystallize in orthorhombic (Pbnm) and rhombohedral structure(R$\overline{3}$c) respectively. The samples are seen to be single phase with the goodness of fit around 1.2 for both cases. For S1 the percentage of Mn$^{3+}$ and Mn$^{4+}$ is respectively about 89\% and 11\% while that for S2 is 74\% and 26\% respectively.  
\maketitle\section{Results and discussions} 
\subsection{Realization of orbital domain state in stoichiometric (S1) sample}
Fig.1 shows the frequency dependence of real part ($\chi$$^R$) of ac susceptibility. The fall in susceptibility at lower temperature is also observed when measurement are done on a single crystals of La${_{0.9}}$Sr$_{0.1}$MnO${_3}$ and La${_{0.875}}$Sr$_{0.125}$MnO${_3}$\cite{sunil}. The temperature variation of $\chi$$^R$ shows paramagnetic to ferromagnetic transition around 175K followed by a hump and the imaginary part ($\chi$$^I$) also shows a peak and a fall at lower temperature. AC-$\chi$ is seen to be frequency dependent below T$_c$ with $\chi$$^R$ decreasing as frequency is increased, normally observed in metastable system like spin glasses, cluster glasses, superparamagnets, re-entrant systems etc. However, there is no shift in temperatures of the peaks with frequency for $\chi$$^R$ and $\chi$$^I$. A frequency dependent peak, which shifts toward higher temperatures with increasing frequency is a characteristic features of the dynamics of spin glass system and is also observed in other manganite samples which shows cluster glass like behaviour\cite{fre}. This observation rules out any spin glass like dynamics, superparamagnetic or cluster glass type behaviour in the low temperature region.

Reports in literature\cite{ark} shows the presence of a low temperature structural transition in this compound.  This results in changed occupancy of orbitals by e$_g$ electron due to change of lattice constant leading to reformation of domains with larger number of domain walls. Sr$^{2+}$ substitution results in inhomogeneous distribution of Mn$^{3+}$ and Mn$^{4+}$ with Mn$^{4+}$ concentration around the divalent ion\cite {shi}, resulting in formation of clusters. These clusters break into smaller pieces due to low temperature structural transition leading to segregation of charge, making the low temperature phase inhomogeneous. The reformation of domains taking place leads to enhance wall number, which are pinned to the new structure. Hence the dynamic response of the spin decreases with decreasing temperature, as the low field is not sufficient to activate the pinned walls resulting in the observed fall in susceptibility. Hence, the resulting self-organized regimes in the form of clusters of various sizes makes the low temperature region of the sample metastable.

To further emphasize the fact that magnetic transition is absent in the low temperature region thermal cycling in both ac and dc magnetisation is done. The presence of thermal hysteresis is a general phenomenon associated with first order phase transition (FOPT). The susceptibility curve both ($\chi$$^R$ and $\chi$$^I$) doesn't show thermal hysteresis around the region where the fall in observed(Fig 2a and its inset). Fig 2b show Field Cooled Cooling (FCC) and Field Cooled Warming (FCW) cycles of DC Magnetisation (DCM). Unlike the Zero Field Cooled Magnetisation (ZFCM) case, the field cooled magnetisation rises continuously with decreasing temperature. The graph also shows absence of thermal hysteresis throughout the temperature range between FCC and FCW. These observations rule out any ferromagnetic to antiferromagnetic FOPT in the low temperature region of the compound. To substantiate the above fact Magneto-Caloric Effect (MCE) measurement is done on the sample (shown in inset of Fig 2b). The entropy change, calculated from MH isotherm at different temperature shows a peak around T$_c$ with absence of any significant peak a lower temperature region. This also indicates absense of FM-AFM transition at lower temperature as a peak in MCE is expected around the transition. Hence the low temperature phase, as seen from thermal hysteresis in magnetisation and MCE measurement is different from the metastable state arising from standard first order transition between competing ferromagnetic and antiferromagnetic phases where near the transition a short-range correlation of one of the two phases start building up at the cost of other. 

Hence the above measurements indicate absence of antiferromagnetic transition, spin glass dynamics or cluster glass like behaviour in the low temperature region, and brings out the novel role of orbitals for explaining the observed features of the sample. The presence of low temperature structural transition in this compound leads to orbital rearrangement, resulting in orbital degree of freedom of e$_g$ playing the central role in defining the ground state properties. So, absence of antiferromagnetic state along with insulating behaviour of the transport which shows a slope change around T$_c$\cite{uru} (due to decrease in the value of resistivity), indicates the coexistence of ferromagnetic metallic and ferromagnetic insulating phase at low temperatures. The insulating behaviour of the ferromagnetic phase is explained in terms of antiferro (AF)-type orbital ordering, which leads to elongation and compression of the neighboring MnO$_6$ octahedrons resulting in unequal Mn-O bond distances. According to Goodenough's theory of semicovalence\cite{goo} the magnetic coupling will be ferromagnetic when the Mn-O bonds are semi covalent leading to ferromagnetic super exchange interaction. Such type of magnetic coupling in similar compounds is also reported in literature \cite{nij,end,kris}. Hence the low temperature phase is an electronically and hence magnetically inhomogeneous state consisting of hole poor and hole rich regions.   So, an orbital domain state with ferromagnetic insulating domain separated by ferromagnetic metallic wall as observed from NMR measurements in La${_{0.8}}$Ca$_{0.2}$MnO${_3}$ \cite{pap} is also realised in our case.

Orbital domains realized in the sample make the low temperature region metastable resulting in a non-equilibrium state where reformation of domains takes place. To probe the energy landscape in this regions degaussing (DG) experiment\cite{joy} is done at different DC fields. Such demagnetisation based studies is considered to give a systematically better approximation of the ground state of disordered systems as reported in Ref\cite{zar}. In DG measurements, after ZFC to 85K, 1000Oe field is applied and then it is reduced to zero. Application of the field disturbs the ground state spin arrangement and results in some remanent magnetization. The remanent magnetization is reduced to zero by repeated field cycling with reducing amplitude (degaussing). Then the measuring field is applied at 85K and temperature response of magnetization is noted while warming. Fig. 3 shows the M-T curves in different measuring fields for degaussed, as well as the corresponding ZFC states. At 9Oe the degaussed curve obtained below the normal ZFC curve while for 30Oe the bifurcation between the curves reduces. At 41Oe the DG curve is above the normal ZFC curve. Again, the seperation between the curves decreases at 55Oe and there is a crossover at 65Oe.At 82Oe the degaussed curve is well below normal ZFC curve. The observed behaviour arise due to the fact that the normal ZFC and degaussed state are different in terms arrangement and size of domains even though the net dipole moment is zero in zero field (before the measuring field is applied). At 85K when a high field is applied it results in formation of large domains, which are broken into smaller pieces by external perturbation (degaussing). Hence the resultant domain size and arrangement are different from that obtained for normal ZFC at 85K.  So when measuring field is applied after ZFC and DG, it lead to different domain size for each case resulting in the observed difference in temperature response of magnetization between them. This behaviour vividly demonstrate inhomogeneous nature of magnetic state, which is not in equilibrium due to reformation of domains. Many metastable configurations are present within which the wall can make thermally activated hops. When the sample is degaussed after a high field was applied, it results in formation of the subvalleys with the moments being locked in certain regions and directions. So measurement at different DC field after degaussing shows different behaviour for each field when compared with normal ZFC measurements indicating a hierarchical organization of energy landscape\cite{lef} which is discussed in details later.
\subsection{Suppression of orbital domains and establishment of ferromagnetic long-range order by non stoichiometry in  
S2 sample}
Fig 4a, shows the temperature dependence of $\chi$$^R$ in different fields of S2 sample. It clearly shows paramagnetic (PM) to ferromagnetic transition (FM) with the absence of any further transition at low temperature. Absence of strong field dependence indicates the presence of long range ferromagnetic ordering where domain wall dynamics in an infinite ferromagnetic matrix plays a significant role in defining physical properties of the compound. More vivid manifestation of the role of the walls is emphasized in the inset of Fig. 4a, which shows the frequency dependence of $\chi$$^R$. Increase in $\chi$$^R$ with the increasing frequency, an intriguing aspect because it is expected $\chi$$^R$ to decrease with the increasing frequency, as observed for metastable systems. In general, the wall distributions for these types of samples is not in equilibrium and are located in position corresponding to the local potential minima around pinning centers and oscillate around these metastable position in response to small AC field. Time dependence measurement of susceptibility performed on LaMnO$_{3.075}$ shows that $\chi$$^R$ decreases with time faster for lower frequencies than at higher frequencies below T$_c$.\cite{mur} This implies walls in a given time stabilize more for a lower frequency than for higher frequency. Moreover the energy of excitation by the of AC field is proportional to the square of its frequency. So higher frequency might provide extra perturbation to the pinned walls for depinning, resulting in larger response of spins with increasing frequency. Hence, the observed field and frequency dependence are quite in contrast to that of S1 where systematic frequency and strong field dependence is observed which is ascribed to the distribution of cluster size with the whole clusters being affected by field and frequency change.

To further highlight the role of the domain wall in S2, thermal hysteresis (TH) in AC-$\chi$ is performed.The PM to FM transition is second order in nature and hence it is expected that TH to be absent. TH is not observed in $\chi$$^R$ (inset of Fig 4b) as it is dominated by the volume response of the domains and is much less sensitive than imaginary part ($\chi$$^I$) to the domain wall dynamics. However, a clear difference is seen in the heating and cooling cycle of $\chi$$^I$ (Fig 4b) which arises out of domain wall motion in the low field regime. The difference in temperature cycle of $\chi$$^I$ (which corresponds to the magnetic losses) indicates thermally irreversible domain wall dynamics due to low field irreversible domain wall pinning in the sample. The TH in $\chi$$^I$ disappears (not shown) in presence of superimposed DC field as the superimposed field is expected to suppress the wall dynamics, emphasizing the above fact that the observed hysteresis is due domain wall motion. 

Degaussing (DG) measurements performed on the S2 sample shows no change in nature of temperature response of magnetization at different DC fields between the normal curve and the curve noted after degaussing (Fig 5), with the DG curve always lying below the normal curve. The difference between the curves (Fig 3 vs. Fig 5) substantiates the fact that the observed features of S2 is only due to the wall dynamics unlike S1 where the domains as a whole is affected by the above protocol.

Hence, even though both the samples show ferromagnetic behaviour, there is a changeover from an orientationly random cluster arrangement of the S1 sample into a homogeneous ferromagnetic ordering for the S2 sample.  Hence it may be considered that S1 sample is constituted of magnetic clusters which are in a metastable state. The interaction among the clusters results in a glassy state which is responsible for non equilibrium nature of the low temperature region. The S2 sample consists strongly coupled regions of equilibrated domains whose once developed correlation are hard to destroy when the temperature is changed. Such behaviour is similar to that of low-doped cuprates where there is a competition between the striped and superconducting phase with the change in oxygen stoichiometry.\cite{cho}  
\subsection{Observation of glassy ferromagnetism in S1 sample and stable ferromagnetism in S2 sample}
As stated earlier, the orbital domain state realized in S1 sample results in segregation of charge making the low temperature region inhomogeneous. To further substantiate the inhomogeneous nature and also to get a better insight about the underlying nature of low temperature magnetic ground state of the compound, time dependent magnetization studies under various heating and cooling protocols have been performed. Fig. 6a shows one such protocol under field cooled cooling and warming condition. Here the temperature response of magnetisation is noted during field cooled cooling (FCC) from room temperature in 9Oe magnetic field with temporary stops at 110K and 95K for a waiting time 7200sec. During the waiting time the field is switched off. After each stop at wait temperatures the field is re-applied and cooling is restarted. Field cooled warming (FCW) curve is noted immediately after the cooling cycle. Decay in magnetization (ageing affect) is observed at the wait temperatures in FCC mode. Instead of memory of aging, significant fluctuation in magnetization obtained in the warming cycle upto 115K. To crosscheck the fluctuation, immediately after the warming cycle the sample is again cooled in 9Oe from room temperature to 85K without any stop and a field cooled warming (FCW) measurement is done. In this case the FCW curve (FCW$_{ref}$ in Fig. 6a) is smooth with absence of any fluctuation indicating that the fluctuation is intrinsic to the sample and is not because of the measuring instrument. It may be noted that in ferromagnetic phase, memory effect is absent during re-heating as it is erased by growth of ferromagnetic domains whereas for a spin glass phase memory of aging can be observed during heating\cite{vin}. Hence absence of memory in our case rules out coexistence of spin glass behaviour with ferromagnetic state (i.e. re-entrant spin glass phase) at lower temperature. In FCC measurement with stopping it is seen that ageing makes the system stiffer with time resulting in lesser response of the spins with field. Fluctuation obtained in the field cooled warming run indicates that domain wall jumps, as the temperature is swept\cite{mur}. Actually the material being inhomogeneous randomly distributed pinning centres prevent the domain wall from establishing the equilibrium position. Hence the above measurements give definite evidence that the low temperature region of the compound is inhomogeneous and is not in a state of global minimum.  

The above ageing measurement performed on S2 sample is shown in Fig 6b. Ageing effect is observed at the wait temperatures 110K and 95K with the effect being more prominent at the higher temperature.This indicates waiting at 110K lead to stabilization of dynamics of the domain walls resulting in lesser prominence of the effect at 95K. During the warming cycle no memory effect of the wait temperatures is observed, as expected in a ferromagnetic phase. Also, absence of magnetic fluctuation in warming cycle indicates the stable nature of the low temperature phase of this sample as compared to that of S1.

To further investigate the effect of ageing, wait time (t$_w$) dependence of ZFC TRM of both the samples is studied.  Fig 7a. shows M vs. t measured with different t$_w$ =1800s, 7200s, 10800s before the application of magnetic field at 95K. As observed, magnetization clearly depends upon the wait time with M value decreasing with the increasing t$_w$ for S1 sample. The behaviour is obvious, as with increasing t$_w$ the system becomes stiffer as if the system sinks in deeper and deeper energy valley as time elapses resulting in lower value of the measured M. In contrasts, even though ageing is observed for the S2 sample M vs. t behaviour is independent of t$_w$ (inset of 7a) as this sample is more ordered than S1. 

For gaining further insight about the underlying nature of the magnetic phase of the samples low field thermoremanent magnetisation measurements are performed. Fig 7b shows the time dependence of magnetisation (TRM) of the S1 sample at different temperature under field cooled (FC) conditions. For each case the sample is cooled from 250K in 9Oe to the measurement temperature where after waiting for 2min the field is switched off and magnetisation decay is noted. Among the various functional form that have been proposed to describe the change of magnetisation with time, the one proposed by Ulrich et al.\cite{ulr} 
\begin{center}
$M(t)= $M$_0$t$^{-\gamma}$....(1)
\end{center}

gave good results of fits, while the other functional form yielded unphysical value of constants with large error bars. In the equation M$_0$ is related to intrinsic ferromagnetic component and exponent ${\gamma}$ depends on strength of magnetic interaction. The values of the parameters for S1 are complied in Table 1. As expected the value of M$_0$ increases with the decreasing temperature as field cooled magnetisation value increases with decreasing temperature but ${\gamma}$ decreases upto 114K and then increases again.
\begin{table}
\caption{\label{tab:table 1}Values of fitting parameters M$_0$ and ${\gamma}$ of equation (1) for the sample S1}
\begin{ruledtabular}
\begin{tabular}{cccc}
T(K) &M(emu/mol)&${\gamma}$(10$^{-3}$)\\
\hline
85 &368.4$\pm$0.15 &8.9$\pm$0.06\\ 
\hline
95 &322.5$\pm$0.1 &4.0$\pm$0.05\\ 
\hline
105 &299.8$\pm$0.1 &3.5$\pm$0.07\\
\hline
114 &287.6$\pm$0.07 &3.4$\pm$0.04\\
\hline
125 &280.2$\pm$0.17 &7.7$\pm$0.01\\
\hline
141 &245.3$\pm$0.15 &8.0$\pm$0.04\\
\end{tabular}
\end{ruledtabular}
\end{table}
Generally for glassy systems the exponent (${\gamma}$) lies between 0 and 1 and also, in our case the value of ${\gamma}$ lying between the mentioned limits indicating a weak intercluster interaction. For spin glasses or a system of interacting particles with fixed size and concentration ${\gamma}$ is expected to be constant with temperature. The variation in the value of ${\gamma}$ as observed is ascribed to the variation of cluster size with temperature indicating the cluster size is very fragile to temperature change.  This indicates a distribution of potential barrier over which the cluster magnetization tends to relax.The value of ${\gamma}$ being lower around 105K-114K is also another signature of the occurrence of orbital rearrangements in the sample. 

For S2 sample, ZFC TRM measurements are done where the field is turned on at the measuring temperature after cooling it from room temperature (Fig 7c). After the field is switched on magnetization shows a sudden increase in value followed by a very slow increase over the measurement time. TRM (normalized with respect to M value at t=0) at different temperature almost superimposes on each other indicating that the relaxation at different temperature is almost the same. Good results are not obtained when the curves are fitted by the available functional form that have been proposed to describe the change of magnetization with time indicating the growth is neither exponential nor logarithmic.  

The nature of the phase in regions where non equilibrium glassy behaviour is observed is generally described either in terms of droplet model\cite{fish} or in terms of hierarchical model. The droplet model introduce the concept of overlap length (L${_{\Delta}}$$_T$), which determine the maximum length scale at which the spin correlation at two different temperatures (the temperatures being less than spin glass transition temperature) are the same. This characteristic length for the group of spins only at distances larger than L${_{\Delta}}$$_T$ is sensitive to small temperature changes. Thus restart of domain growth is observed from the size L${_{\Delta}}$$_T$ not only after cooling but also after heating implying a symmetrical behaviour with respect to positive/negative temperature cycle. In this model it is believed that, at any given temperature below spin glass transition, there is only one phase (and its spin reversed counterpart) to be considered. Hence it can be said, that the energy landscape in this case is dominated by one large valley unlike for hierarchical model, where a multi valley structure is hierarchically organized on the free energy surface. Here the free energy landscape consists of many local minima corresponding to metastable configuration, which splits into new state when temperature decreases and merges back when the temperature is raised. Hence hierarchical picture predicts that relaxation is fully initialized on heating implying a non symmetrical behaviour with respect to heating and cooling unlike the droplet model. Hence a series of TRM measurement with temperature change as proposed by Sun et al.\cite{sun} is performed to associate the energy distribution at low temperature phase of the samples with one of the above defined model. The relaxation measurement for sample S1 is shown in Fig 8a. The sample is cooled from 240K to 95K in 0Oe/9Oe field. At 95K after waiting for 120sec the field was switched on/off and magnetisation is noted for time t1 = 1hour. The sample was then cooled in constant/ zero field to 85K and TRM is measure for time t2 = 1hour. Then the sample was heated back to 95K in constant/zero field and TRM was measured for time t3 = 1hour. For the ZFC case, during t1 the curve shows an immediate rise followed by steady growth after the field is switched on. During temporary cooling the relaxation is weak. Again when the temperature is raised to 95K the magnetisation start from the value it reached at the end of t2 indicating absence of reinitialization after the cooling cycle. For the FC case, magnetisation shows an immediate fall followed by steady decay after the field is switched off. During temporary cooling magnetisation start from a higher value but the relaxation is weak. During t3 the relaxation curve starts from a level which is near to the value reached at the end of t1. Fig 8b shows the above protocol in the heating cycle where the relaxation curves are noted at 95K, 105K and 95K for time t1, t2 and t3 (one hour each) respectively. Every time the starting value of magnetisation is different from the value it reached at the end of previous TRM measurement. Hence, a clear reinitialization in the relaxation is observed during temporary heating in both ZFC and FC cases. Therefore it can be said that there is an antisymmetric response with respect to positive and negative temperature change in TRM measurement in both ZFC and FC process which favours a hierarchical picture of energy landscape in the low temperature region which have also been suggested in the earlier section. Interestingly, such picture of energy landscape has also been proposed for many compound like interacting magnetic nanoparticle system\cite{sun}, re-entrant systems\cite{vin} etc. The collective interactions of the self-generated assembly of clusters in the low temperature ferromagnetically inhomogeneous phase in our case may give rise to a glassy magnetic behaviour which constitute a new class of glass different from conventional spin glass as also reported in Ref \cite{riv}.

The above procedure is performed for sample S2 by cooling it from room temperature as shown in Fig 9a and 9b. In this case the change of magnetisation with time is very small and the observed minor change in relaxation behaviour during t2, t3 is only due to change in magnetisation value with temperature change. So the energy landscape of magnetic phase of S2 cannot be ascribed to any of the above models. 
\maketitle\section{Conclusion}
In summary, we have tried to solve the controversy related to the magnetic ground state of in low doped manganite systems through bulk magnetic measurements on La${_{0.9}}$Sr$_{0.1}$MnO${_3}$. Such systems show a well defined paramagnetic-ferromagnetic transition with the decrease in temperature which falls into the isotropic 3D Heisenberg universality class. However, with further decrease in temperature there is a sharp change in magnetic susceptibility which is attributed to inhomogeneous ferromagnetism. This inhomogeneity is considered to be arising from the formation of orbital domain state (comprising of ferromagnetic insulating domains separated by ferromagnetic metallic walls) resulting from a discontineous change of lattice parameters at low temperature. This self organised regimes show metastability which is different from that arising from broad first order phase transitions. It is clearly shown that the low temperature phase shows glassy behaviour which is different from conventional spin glass, cluster glass or dynamics observed in reentrant systems. This glassy phase shows ageing affect but no memory and the energy landscape of the degenerate ground state follows the picture of hierarchical model.

To conclusively assert the fact that the orbital degrees of freedom of the e$_g$-electrons plays an important role in defining the ground state of the system, non-stoichiometry is introduced. Disorder in form of self doping reduces the Mn$^{3+}$ and hence e$_g$-electrons by 17\% in the La${_{0.9}}$Sr$_{0.1}$MnO$_{3.08}$ sample. This leads to complete destruction of orbital domain state of the stoichiometric sample resulting in homogeneous ferromagnetic ordering. The ferromagnetic phase of this non-stoichiometric sample does not show glassy behaviour and the energy landscape picture of the sample is neither in accordance with hierarchical model or droplet model. Further studies on stoichiometric sample in terms of coupling of spin, orbitals with lattice degrees of freedom and their dynamics will be useful in understanding the observed unusual glassy behaviour of the system. These studies will be important in establishing analogy between self-organised regimes of low doped manganites with that of cuprates and nickalates.

\maketitle\section{Acknowledgement}
We are grateful to Dr. P. Chaddah for many fruitful discussions. We are thankful to Mr. K. Kumar and Mr. A. K. Pramanik for help rendered during the course of measurement. KM acknowledges CSIR, India for financial support. 

\newpage

\begin{figure}
	\centering
		\includegraphics{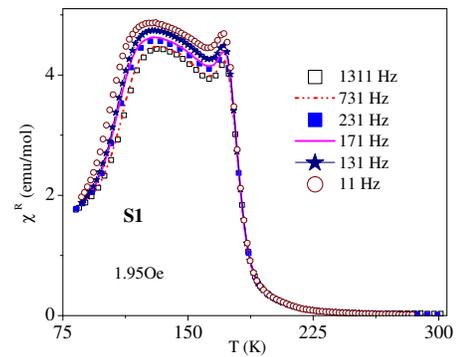}
	\caption{(color online) Frequency dependence of real part of ac susceptibility ($\chi$$^R$) of S1 sample}
	\label{fig:Fig1}
\end{figure}

\begin{figure}
	\centering
		\includegraphics{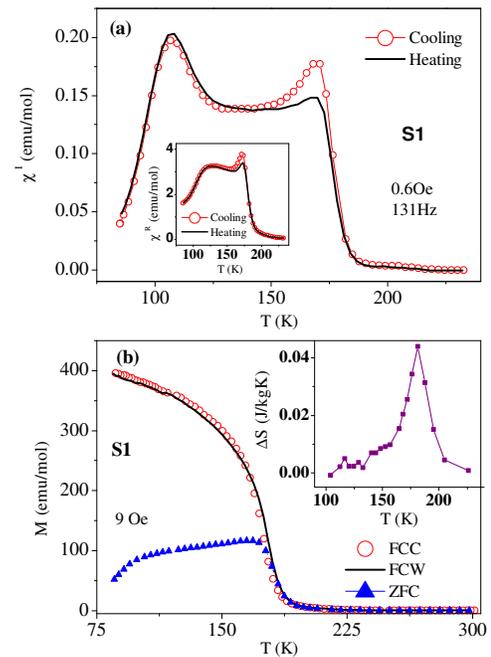}
	\caption{(color online) (a) Thermal hysteresis (TH) imaginary part of ac susceptibility ($\chi$$^I$) of S1 sample. Inset shows TH of the real part of ac susceptibility ($\chi$$^R$). (b) Temperature response of field cooled warming (FCW)  and field cooled cooling (FCC) curves of  DC magnetization along with zero field cooled magnetisation (ZFCM) curve at 9Oe of S1 sample. Inset shows Temperature dependence of Magnetocaloric effect of the same.}
	\label{fig:Fig2}
\end{figure}

\begin{figure}
	\centering
		\includegraphics{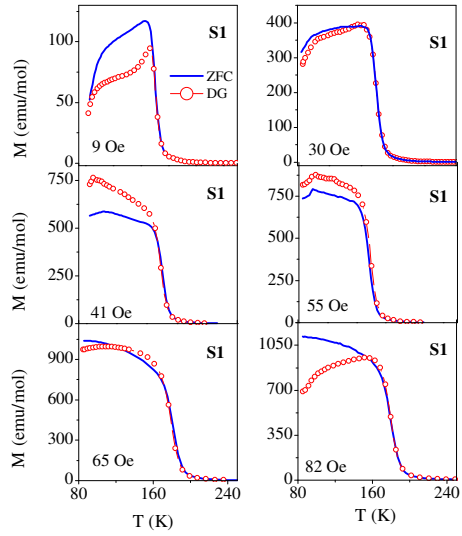}
	\caption{(color online) Temperature response of magnetization after zero field cooling (ZFC) [line] and applying a field and degaussing the ZFC sample (DG) [open circles] at different magnetic fields for S1 sample.}
	\label{fig:Fig3}
\end{figure}

\begin{figure}
	\centering
		\includegraphics{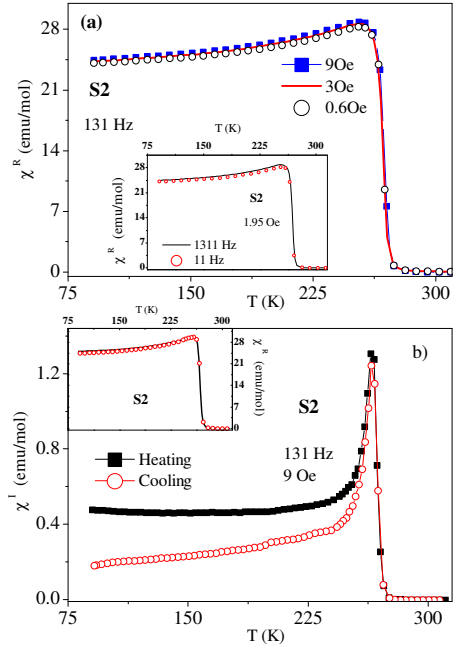}
	\caption{(color online) (a) Field dependence of real part of ac-susceptibility ($\chi$$^R$) of  S2 sample. Inset shows frequency dependence of real part of ac susceptibility ($\chi$$^R$)  of the same.  (b) Thermal hysteresis (TH)  of the imaginary part of the ac susceptibility ($\chi$$^I$) for S2 sample. Inset shows the TH of real part of ac-susceptibility ($\chi$$^R$).}
	\label{fig:Fig4}
\end{figure}

\begin{figure}
	\centering
		\includegraphics{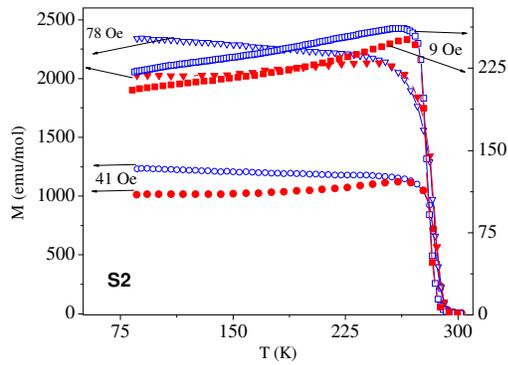}
	\caption{(color online) Temperature response of magnetization after zero field cooling (ZFC) [open symbols] and applying a field and degaussing the ZFC sample [closed symbols] at different magnetic fields for S2 sample.}
	\label{fig:Fig5}
\end{figure}

\begin{figure}
	\centering
		\includegraphics{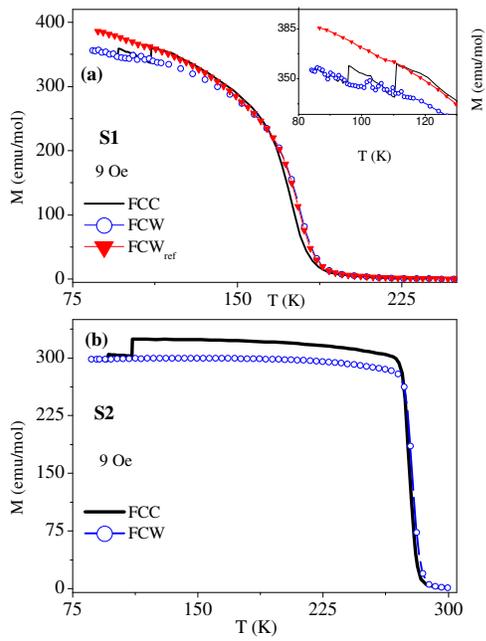}
	\caption{(color online) (a) M-T curves during field cooling. The field is switched off at two temperatures (110K and 95K) for a waiting time of 7200 s. The M-T curve in warming mode and normal FCW curve (as FCW$_{ref}$ ) is also shown. Inset shows the above graphs upto 125K. (b)  Above protocol (only FCC and FCW) for S2 sample.}
	\label{fig:Fig6}
\end{figure}

\begin{figure}
	\centering
		\includegraphics{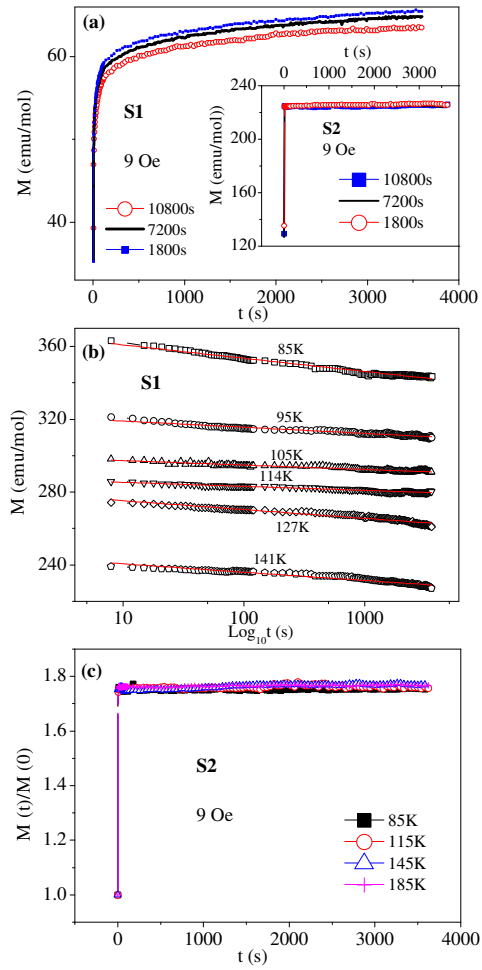}
	\caption{(color online) (a) M vs. t plot of S1 sample at 95K for different wait times (t$_w$). Inset shows the same measurement for the S2 sample. (b) M is plotted against log t (in sec) for S1 sample at different temperatures. The solid lines are best fit to Eq. (1). (c) Normalized magnetic moment M(t)/M(0)  is plotted  against time for S2 sample after zero field cooling to the measurement temperature and switching on the field.}
	\label{fig:Fig7}
\end{figure}

\begin{figure}
	\centering
		\includegraphics{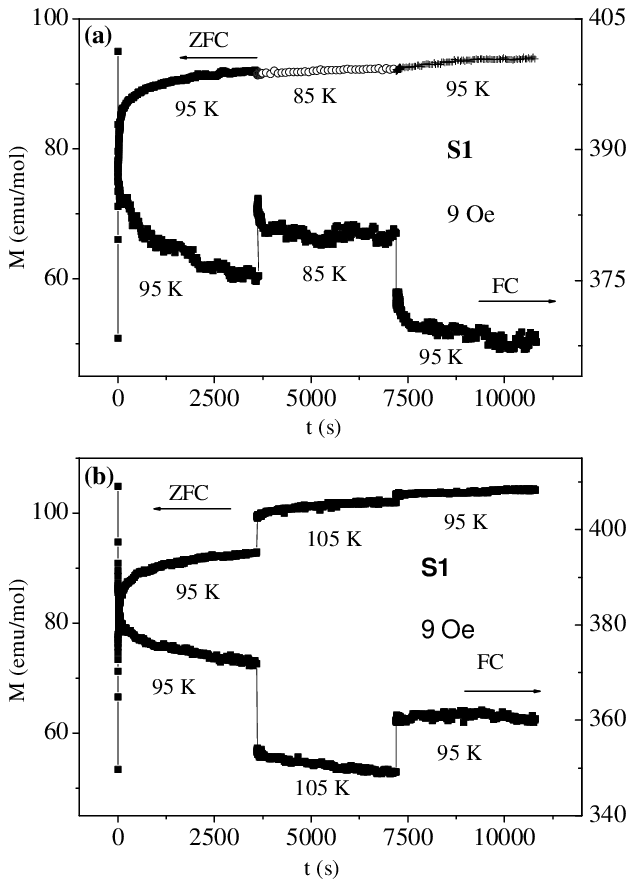}
	\caption{(a) Magnetic relaxation with temporary cooling for both ZFC (t1, t2,t3 for one hour each) and FC (t1, t2, t3) for one hour each) cases for S1 sample (b) Magnetic relaxation with temporary heating for both ZFC (t1, t2, t3 for one hour each) and FC (t1, t2, t3 for one hour each) cases for S1 sample}
	\label{fig:Fig8}
\end{figure}

\begin{figure}
	\centering
		\includegraphics{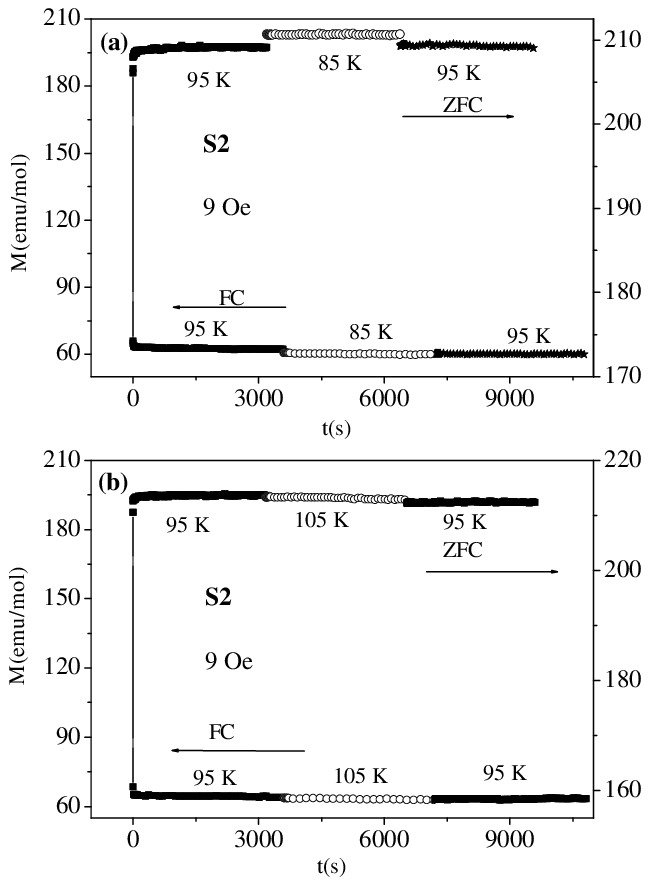}
	\caption{(a) Magnetic relaxation with temporary cooling for both ZFC (t1, t2, t3 for 50 min each) and FC (t1, t2, t3 for one hour each) cases for S2 sample. (b) Magnetic relaxation with temporary heating for both ZFC (t1, t2, t3 for 50 min each) and FC (t1, t2, t3 for one hour each) cases for S2 sample.}
	\label{fig:Fig9}
\end{figure}

\end{document}